# Molecular Dysprosium Complexes for White-Light and Near-Infrared Emission Controlled by the Coordination Environment

Dimitrije Mara[a*], Flavia Artizzu[b], Joydeb Goura[b], Manjari Jayendran[b], Bojana Bokic[c], Branko Kolaric[c,d], Thierry Verbiest[a], Rik Van Deun[b*]

[a] – Molecular Imaging and Photonics, Department of Chemistry, KU Leuven, Celestijnenlaan 200 D, box 2425, B-3001, Leuven, Belgium.
[b] – L$^3$ – Luminescent Lanthanide Lab, Department of Chemistry, Ghent University, Krijgslaan 281 – S3, B-9000, Ghent, Belgium.
[c] – Photonics Center, Institute of Physics, University of Belgrade, Pregrevica 118, 11080, Belgrade, Serbia.
[d] – Micro- and Nano-photonics Materials Group, University of Mons, Avenue Maistriau 19, B-7000, Mons, Belgium.

[*] Corresponding authors: dimitrije.mara@kuleuven.be (D. Mara), rik.vandeun@ugent.be (R. Van Deun)

**Abstract**

A series of single-molecule dysprosium ($Dy^{3+}$) complexes consisting of β-diketonate ligands, $L_1$ = 4,4,4-trifluoro-1-phenyl-1,3-butadionate and $L_2$ = 4,4,4-trifluoro-1-(4-chlorophenyl)-1,3-butadionate, as water-containing complexes, and the auxiliary triphenylphosphine oxide (tppo) ligand as water-free complexes were investigated as potential white-light emitters. The coordination environment and choice of the ligands play an important role in the behavior of the yellow/blue emission of the $Dy^{3+}$ complexes (Y: $^4F_{9/2} \rightarrow {}^6H_{13/2}$ – yellow, and B: $^4F_{9/2} \rightarrow {}^6H_{15/2}$ and ligand phosphorescence – blue) based on the sensitization efficiency of the $Dy^{3+}$ ion by the ligands. By introducing the auxiliary tppo ligand in the complex, the relative intensity of the $Dy^{3+}$ emission increases due to a more efficient sensitization of the $Dy^{3+}$ ion. The CIE (Commission International d'Eclairage) coordination at room temperature for water-containing, **DyL$_1$H$_2$O** (0.340, 0.333), and **DyL$_2$H$_2$O** (0.270, 0.249), and for water-free complexes, **DyL$_1$tppo** (0.364,0.391) and **DyL$_2$tppo** (0.316, 0.331), are close to the coordinates of 'ideal' white light (0.333, 0.333). The CCT (Correlated Color Temperature) values at room temperature for **DyL$_1$H$_2$O** (5129 K), **DyL$_2$H$_2$O** (18173 K), and **DyL$_2$tppo** (6319 K) correspond to 'cold-white-light' emitters, while the **DyL$_1$tppo** (4537 K) matches a 'warm-white-light' emitter. Beside emitting in the visible (Vis) region, the $Dy^{3+}$ complexes also show emission in the near-infrared (NIR) part of spectrum, which has been studied in detail.

Keywords: White-light emission, Lanthanide complexes, Luminescence, NIR emission

## 1. **Introduction**

The growth of the global energy consumption has accelerated development and usage of energy-saving smart devices and the energy-efficient solid-state lighting (SSL). Solid-state white-light-emitting materials posses exceptional properties such as energy saving and long operational lifetime, which has already led to their widespread application covering large-panel displays to ambient lighting. [1-3] First, the SSL sources can be subdivided based on the materials, from which they are made of, either inorganic phosphors (LED) or organic molecule semiconductors (OLDE). From another perspective, the way they are stimulated to emit light, the SSLs are subdivided in two categories: UV excitation (LED) or by electrical excitation (OLED). SSLs are much more efficient compared to the classical incandescent lamps, or environmentally friendly as compared to fluorescent and mercury lamps. [4-6]

In general, there are two different approaches to obtain white-light emission (WLE): using dichromatic emitters (blue and yellow (B/Y) or blue and red (B/R)), or by using trochromatic emitters which combine three primary colors (red, green, and blue (RGB)). [1,7-19] White light is obtained by using separate dopants or multiple phase matrices such as organic compounds, metal complexes, nanomaterials, hybrid organic-inorganic materials or inorganic phosphors. All these types of materials have some unique characteristics to exploit their advantages in obtaining white light, but sometimes it is necessary to mix different types of materials to achieve the intended goal. On the other hand, white-light emission (WLE) from single molecular materials can also achieved, spanning the whole visible (Vis) spectrum. [20-25] The advantage of single molecular materials is that they consist of a single emitter, and this simplifies the production of lighter and thinner materials, which is highly favorable. [26]

Additionally, lanthanide complexes, because of the unique properties of the lanthanide ($Ln^{3+}$) ions, which have electrons in 4f orbitals, are interesting platforms for the applications of those materials in lighting technologies. The $Ln^{3+}$ luminescence arises from the 4f – 4f transitions, which correspond to characteristic colors (wavelengths) from the ultra-violet (UV) across the Vis to the near-infrared (NIR) spectral region. Unfortunately, the 4f – 4f transitions are difficult to excite directly according to Laporte's rule and a way to overcome this disadvantage is by using organic chromophores which act as antenna to sensitize the $Ln^{3+}$ ions. [27] β-Diketonate complexes are well known in lanthanide coordination chemistry and have been studied in extent. [28,29] Their ability to excite virtually every spectroscopically active $Ln^{3+}$ ion to obtain the pure $Ln^{3+}$ luminescence colors, as well as their processability into more complex matrices creates the opportunity for a wide range of applications, such as lighting, ion sensing, temperature sensing, telecommunications, etc. [30-33] The organic chromophores in this case should be designed or chosen to act as an antenna for the $Ln^{3+}$, but at the same time be able to emit by itself. Designing of such WLE $Ln^{3+}$ complexes can be achieved by combining an organic chromophore which emits in the blue region, with a $Ln^{3+}$ ion that emits in the yellow ($Dy^{3+}$) or in the red ($Eu^{3+}$) region. [34-49]

Here, we present a series of single-molecule WLE tris $Dy^{3+}$ β-diketonate complexes of two different β-diketonate ligands (4,4,4-trifluoro-1-phenyl-1,3-butadionate and 4,4,4-trifluoro-1-(4-chlorophenyl)-1,3-butadionate), with either coordinated water molecules or triphenylphosphine oxide (tppo) as co-ligand, which can be excited at 365 nm. [50,51] Simultaneous emission from the ligand and the $Dy^{3+}$ has been observed, giving rise to WLE, which by modifying the coordination environment, could be altered from deep cold (blue) to warm (yellow) white light. In addition to white emission in the Vis region, the series of $Dy^{3+}$ complexes also showed emission in the NIR spectral region.

## 2. **Experimental Section**

### 2.1. Materials

$DyCl_3 \cdot 6H_2O$ (99.9%), 4,4,4-trifluoro-1-phenyl-1,3-butadione (Hbfa) 99% and triphenylphosphine oxide 98% were purchased from Sigma Aldrich. The 4,4,4-trifluoro-1-(4-chlorophenyl)-1,3-butadione 98% was purchased from TCI Europe. Methanol (laboratory grade, 100%) and NaOH were purchased from Fisher Scientific. All chemicals were used without further purification. All reaction were carried out under atmospheric conditions.

### 2.2. Synthesis of $[Dy(L_{1(2)})_3(H_2O)_2]$ and $[Dy(L_{1(2)})_3(tppo)_2]$ complexes

The synthesis procedure has been previously reported in detailed and will be only discussed in short. [50] The synthesis of $[Dy(L_{1(2)})_3(H_2O)_2]$ was done in methanol by first dissolving an appropriate amount of ligand $L_1$ and $L_2$ (0.9 mmol), which were then deprotonated with an equimolar amount of NaOH prior to the addition of methanol solution of $DyCl_3 \cdot 6H_2O$ (0.3 mmol). The obtained crystals were recrystallized from methanol solution and used for further analysis. The synthesis of $[Dy(L_{1(2)})_3(tppo)_2]$ was done in methanol, by addition of a methanol solution of $[Dy(L_{1(2)})_3(H_2O)_2]$ (0.1 mmol) to dissolve triphenylphosphine oxide (tppo) (0.2 mmol) and the complexes were used without any additional purification. No crystals suitable for single crystal X-ray diffraction could be obtained due to the formation of twinned crystals during crystallization.

$[Dy(L_1)_3(H_2O)_2]$ **$DyL_1H_2O$**: Elemental analysis (%) calculated for $C_{30}H_{22}F_9O_8Dy$ (847.00): C 42.54, H 2.98, found: C 42.45, H 2.93. FT-IR (KBr) $\nu_{max}$ ($cm^{-1}$): ): 3657 (s; νst O-H, free),

3449 (w; νst O-H, H-bonded), 3083, 2774, 2601, 2525 (w; νst C-H and Fermi resonance), 2482, 2381, 2321, 2273, 2229, 2137 (w; aromatic overtone), 2085 2044, 1985, 1908, 1863, 1820, 1775 (w; comb, aromatic), 1667 (w; νst C=O, keto form), 1628, 1571, 1532, 1493, 1467 (w; νst ar C-C), 1379, 1288 (w; νst C-F, CF3), 1249, 1197, 1144 (w; νst C-F, CF3 and δip ar C-H), 1096 (w), 1027 (s; δip ar C-H), 996 (s; δoop C-H), 817 (w; γ ar C-C and νst C-Cl), 773, 721 (w; δ C-F, CF3, δoop C-H and γ ar C-C and νst C-Cl). ESI-MS (negative mod, -), m/z: 868.00 $[M+Na-2H]^-$, 1023.97 $[M+2Br+H]^-$ Isotope used for calculation is $^{164}$Dy.

$[Dy(L_2)_3(H_2O)_2]$ **$DyL_2H_2O$**: Elemental analysis (%) calculated for $C_{30}H_{19}Cl_3F_9O_8Dy$ (950.33): C 37.92, H 2.33; found: C 37.88, H 2.25. FT-IR (KBr) $\nu_{max}$ ($cm^{-1}$): 3657 (s; $\nu_{st}$ O-H, free), 3434 (s; $\nu_{st}$ O-H, H-bonded), 3072, 2779, 2677, 2591 (w; $\nu_{st}$ C-H and Fermi resonance), 2508, 2434, 2389, 2327, 2286, 2232, 2133 (w; aromatic overtone), 2096, 2051, 1911, 1796 (w; comb, aromatic), 1681 (w; $\nu_{st}$ C=O, keto form), 1627, 1574, 1536, 1487, 1463 (w; $\nu_{st}$ ar C-C), 1401, 1359, 1294 (w; $\nu_{st}$ C-F, $CF_3$), 1248, 1191, 1146 (w; $\nu_{st}$ C-F, $CF_3$ and $\delta_{ip}$ ar C-H), 1096 (w), 1014 (s; $\delta_{ip}$ ar C-H), 944 (s; $\delta_{oop}$ C-H), 849 (w; γ ar C-C and $\nu_{st}$ C-Cl), 799, 738, 705, 664 (w; δ C-F, $CF_3$, $\delta_{oop}$ C-H and γ ar C-C and $\nu_{st}$ C-Cl). ESI-MS (negative mod, -), m/z: 972.00 $[M+Na-2H]^-$, 1159.00 $[M+2Br+H]^-$. Isotope used for calculation is $^{164}$Dy.

$[Dy(L_1)_3(tppo)_2]$ **$DyL_1tppo$**: Elemental analysis (%) calculated for $C_{66}H_{48}F_9O_8P_2Dy$ (1367.54): C 57.97, H 3.76; found: C 57.88, H 3.70. FT-IR (KBr) $\nu_{max}$ ($cm^{-1}$): 3267, 3147 (w; νst C-O, enol beyond the range), 3061 (s; νst ar C-H), 2965, 2913, 2771, 2710, 2599 (w; νst C-H and Fermi resonance), 2520, 2471, 2434, 2381, 2323, 2278, 2129 (w; aromatic overtone), 2088, 2030, 1969, 1895, 1820 (w; comb, aromatic), 1776, 1713 (w; νst C=O, keto form), 1656, 1577, 1535, 1491, 1442 (w; νst ar C-C), 1376, 1323 (w; νst C-F, CF3), 1290 (w; νst P=O), 1244, 1199, 1146 (w; νst C-F, CF3, δip ar C-H and νst R3P=O), 1094, 1075, 1026 (s; δip ar C-H and νst R3P=O), 997, 972 (w; νst R3P=O), 943 (s; δoop C-H and νst R3P=O), 848, 807, 795 (w; γ ar C-C and νst P-C), 758, 725, 635 (w; δ C-F, CF3, δoop C-H and γ ar C-C). ESI-MS (positive mode +), m/z: 1428.00 $[M+Na+ACN]^+$, 1150.00 $[M\text{-}tppo+Na+ACN]^+$, 870.00 $[M\text{-}2tppo+Na]^+$. Isotope used for calculation is $^{164}$Dy.

$[Dy(L_2)_3(tppo)_2]$ **$DyL_2tppo$**: Elemental analysis (%) calculated for $C_{66}H_{45}Cl_3F_9O_8P_2Dy$ (1470.87): C 53.89, H 3.29; found: C 53.80, H 3.25. FT-IR (KBr) $\nu_{max}$ ($cm^{-1}$): 3265, 3146 (w; νst C-O, enol beyond the range), 3060 (s; νst ar C-H), 2961, 2920, 2829, 2768, 2714, 2586 (w; νst C-H and Fermi resonance), 2467, 2380, 2327, 2286, 2249, 2129 (w; aromatic overtone), 2088, 2039, 1964, 1898, 1820 (w; comb, aromatic), 1776 (w; νst C=O, keto form), 1627, 1594,

1533, 1483, 1438 (w; νst ar C-C), 1380, 1318 (w; νst C-F, CF3), 1288 (w; νst P=O), 1240, 1187, 1141 (w; νst C-F, CF3, δip ar C-H and νst R3P=O), 1088, 1067, 1030 (s; δip ar C-H and νst R3P=O), 1014, 972 (w; νst R3P=O), 930 (s; δoop C-H and νst R3P=O), 849, 783 (w; γ ar C-C and νst P-C), 725, 693, 660 (w; δ C-F, CF3, δoop C-H and γ ar C-C). ESI-MS (positive mode +), m/z: 1497.87 $[M+Na+ACN]^+$, 1218.87 $[M\text{-tppo}+Na+ACN]^+$, 980.00 $[M\text{-}2\text{tppo}+Na]^+$. Isotope used for calculation is $^{164}Dy$.

2.3. Characterization

Luminescence measurements were performed on an Edinburgh Instruments FLSP920 UV-Vis-NIR spectrometer setup. A 450 W Xe lamp was used as steady-state excitation source. Time-resolved measurements were recorded using a 60 W Xe lamp operating at frequency of 100 Hz. A Hamamatsu R928P photomultiplier tube was used to detect emission signal in the visible region. A Hamamatsu R5509-72 multiplier tube was used to detect emission in the NIR region. The absolute quantum yield (QY) of the complex was determined using an integrating sphere coated with BENFLEC (provided by Edinburgh Instruments) and calculated using **Equation 1**:

$$\eta = \frac{\int L_{emission}}{\int E_{blank} - \int E_{sample}} \quad (1)$$

Where $L_{emission}$ is the integrated are under the emission spectrum, $E_{blank}$ is integrated are under the "excitation" band of the blank, and $E_{sample}$ is the integrated are under excitation band of the sample (as the samples absorbs part of the light, this area will be smaller than $E_{blank}$). All luminescent measurements were recorded at room temperature. Crystals were put between quartz plates (Starna cuvettes for powder samples, type 20/C/Q/0.2). Fourier Transform Infrared (FTIR) spectra were acquired in the region of 400-4000 $cm^{-1}$ with a Thermo Scientific Nicolet 6700 FT-IR spectrometer equipped with a nitrogen-cooled Mercury Cadmium Telluride (MCT) detector and KBr beam splitter; samples were measured in KBr pellets. Elemental analysis (C, H, N) was performed on a Thermo Fisher 2000 elemental analyzer, using $V_2O_5$ as catalyst. ESI-MS was performed on an Agilent 6230 time-of-flight mass spectrometer (TOF-MS) equipped with Jetstream ESI source and positive and negative ionization modes were used.

3. Results and discussion

3.1. Synthesis and characterization of $Dy^{3+}$ complexes

The complexes were synthesized using mild reaction conditions, where the β-diketonate ligand was deprotonated with an equimolar amount of sodium hydroxide and reacted with the $Dy^{3+}$ ion in a stoichiometric ratio in methanol. The high reactivity of the β-diketonate ligand toward to the coordination of $Dy^{3+}$ ions prevents the formation of highly insoluble lanthanide hydroxides, which could be formed under basic conditions. The complexes with formula $[Dy(L_1)_3(H_2O)_2]$ ($L_1$ = 4,4,4-trifluoro-1-phenyl-1,3-butadionate) and $[Dy(L_2)_3(H_2O)_2]$ ($L_2$ = 4,4,4-trifluoro-1-(4-chlorophenyl)-1,3-butadionate) were isolated. The synthesis of water-free complexes was done by replacing the water molecules from the first coordination sphere with the triphenylphosphine oxide (tppo) ligand. This was done by introducing tppo ligand in methanol solution in a stochiometric ratio to the $[Dy(L_{1(2)})_3(H_2O)_2]$ complex to form and isolate the water-free complexes with formula $[Dy(L_1)_3(tppo)_2]$ and $[Dy(L_2)_3(tppo)_2]$. These complexes have been characterized by FT-IR, ESI-MS (see SI Figs. S1-S8) and elemental analysis and confirmed that all the complexes have the same compositions as the ones reported previously. [50,51]

## 3.2. Steady-state and time-resolved photoluminescence (PL) studies

### 3.2.1. Steady-state and time-resolved PL of $Dy^{3+}$ complexes in the visible region

Upon excitation into the β-diketonate ligand absorption band (see SI Figs. S9-S12) with UV light all studied $Dy^{3+}$ complexes showed emission in the Vis and NIR spectral range displaying the characteristic $Dy^{3+}$ peaks. In Fig. 1, the PL emission spectra of **$DyL_1H_2O$**, **$DyL_1tppo$** (a), **$DyL_2H_2O$** and **$DyL_2tppo$** (c) are presented. The $Dy^{3+}$ $^4F_{9/2} \rightarrow {}^6H_{15/2}$ transition appearing at 480 nm, the $^4F_{9/2} \rightarrow {}^6H_{13/2}$ transition at 575 nm and the $^4F_{9/2} \rightarrow {}^6H_{11/2}$ transition at 660 nm are clearly observed. Besides the emission peaks of the $Dy^{3+}$ ion, a broad band (400 to 450 nm) with a peak maximum at ~ 430 nm is observed, which is assigned to the emission from the ligand. In water-free complexes the relative intensity of this broad band is lower as compared to its intensity in water-containing complexes. The advantage of the introduction of the tppo ligand into the water-containing complexes brings two benefits: first, the exclusion of water molecules from the first coordination sphere around the $Dy^{3+}$ ion (reducing quenching) and second, a more efficient energy transfer from the tppo ligand to $Dy^{3+}$ ion, increasing the relative intensity of $Dy^{3+}$ peaks in the emission spectra comparison to the ligand band (Fig. 1. a and c)

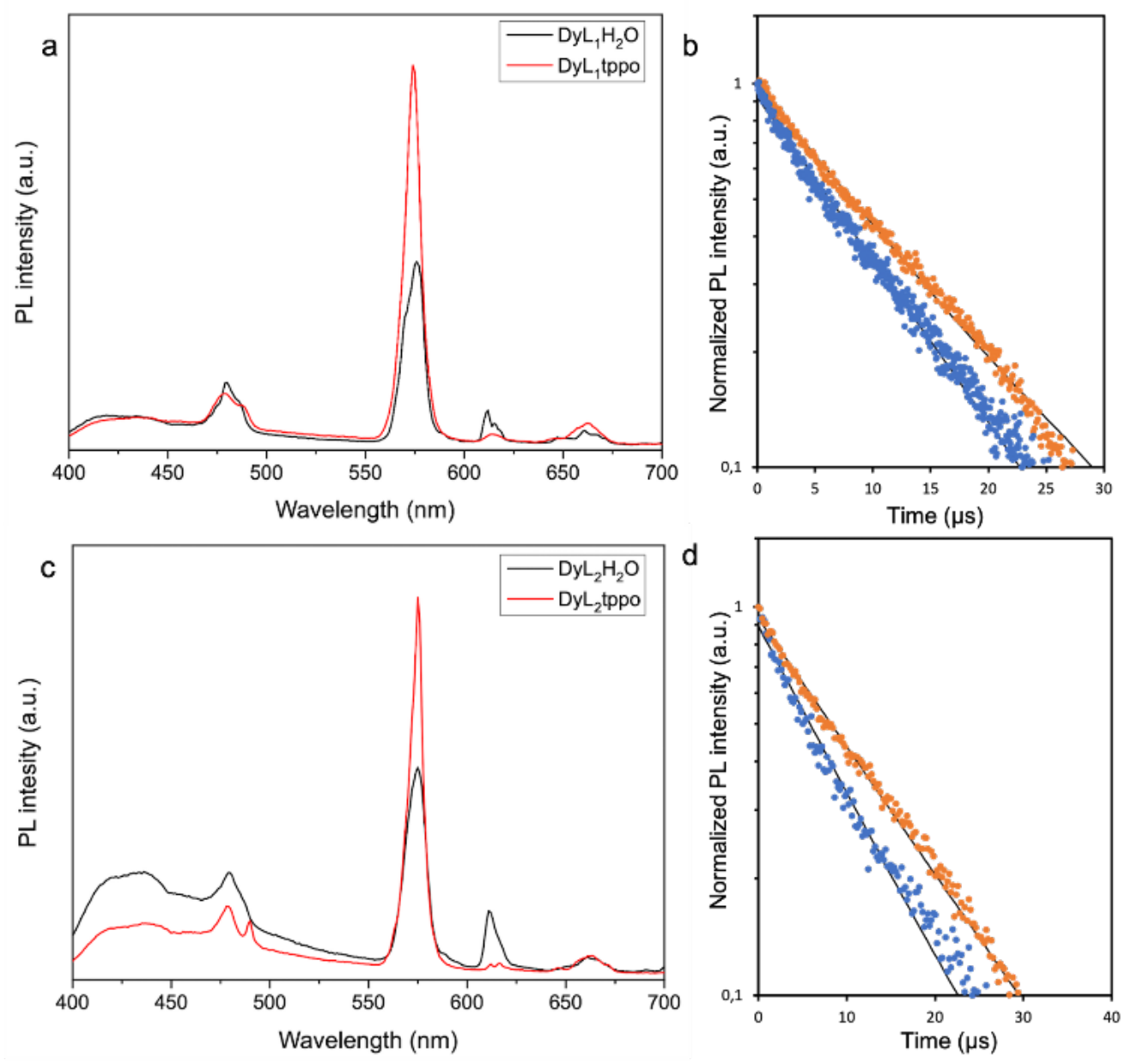

**Fig. 1.** (a) PL emission spectra of $DyL_1H_2O$ and $DyL_1tppo$; (b) decay profile of $DyL_1H_2O$ (blue) and $DyL_1tppo$ (orange) observed at 575 nm; (c) PL emission spectra of $DyL_2H_2O$ and $DyL_2tppo$; (d) decay profile of $DyL_2H_2O$ (blue) and $DyL_2tppo$ (orange) observed at 575 nm. * Contamination of the $DyCl_3{\cdot}6H_2O$ salt with europium salts not influencing the photophysics of $Dy^{3+}$.

The time-resolved measurements show (see Table 1.) that the water-containing complexes have shorter luminescent lifetimes compared to water-free complexes. The sensitization efficiency of the $Dy^{3+}$ ion, in the water-free complexes is estimated to be a slightly higher as well, because of higher triplet state ($T_1$ = 23800 $cm^{-1}$) of the tppo co-ligand compared to the triplet states of ligands $L_1$ ($T_1$ = 22500 $cm^{-1}$) and $L_2$ ($T_1$ = 21500 $cm^{-1}$). [53] The decay profiles for both water-containing and water-free complexes were fitted with monoexponential functions. The differences in the contribution of the neutral tppo ligand between the complexes with ligands $L_1$ and $L_2$ are due to the energy difference between their triplet states ($T_1$). As can be seen in Fig. 2. the energy difference between the triplet state of ligand $L_1$ and emitting level of $Dy^{3+}$ is $\Delta E$ = 1500 $cm^{-1}$, and for the ligand $L_2$ this difference is $\Delta E$ = 500 $cm^{-1}$. Incorporation of the neutral tppo ligand the resulting water-free complexes show slightly longer lifetimes. This is because of the reduced quenching efficiency by high-energy oscillators related to the water molecules in the first coordination sphere.

**Table 1.** Luminescence lifetimes and absolute quantum yields for the $Dy^{3+}$ complexes.

| Compound | τ (μs) | $R^2$ | QY [%] |
|---|---|---|---|
| $DyL_1H_2O$ | 10.24 | 0.995 | 4.94 |
| $DyL_1tppo$ | 11.50 | 0.995 | 5.32 |
| $DyL_2H_2O$ | 9.95 | 0.991 | 3.00 |
| $DyL_2tppo$ | 13.15 | 0.994 | 3.58 |

To have an efficient energy transfer (ET), the difference between the energy donor, being the triplet state ($T_1$) of the ligand, and the acceptor, being the emitting level of $Ln^{3+}$ ion, should ideally be between 2500 and 3500 $cm^{-1}$. An energy difference outside of this range (2500-3500 $cm^{-1}$) decreases the ET efficiency. First, when the energy difference is larger than 3500 $cm^{-1}$, the ET efficiency is lower due to the possibility of nonradiative relaxation (NR) pathways of the ligand, or a loss of energy during the ET process, resulting in an incomplete ET from the ligand to the $Ln^{3+}$ ion. Second, when the energy difference between the $T_1$ level of the ligand and the emitting level of the $Ln^{3+}$ ion is less than 2500 $cm^{-1}$ there is a competition between the ET and back energy transfer (BET) process, which will reduce the efficiency of ET from the ligand to $Ln^{3+}$ ion. [54] In this case, when the triplet states of ligands $L_1$ and $L_2$ (donors) are close to the emitting level of $Dy^{3+}$ (acceptor), the BET process is much more competitive with the ET process (see Fig. 2.) Ligand emission is observed in both water-containing complexes. The relative intensity of the $L_2$ ligand emission is stronger than that of the $L_1$ emission. The difference in intensity is a result in more dominant BET as compared to the ET.

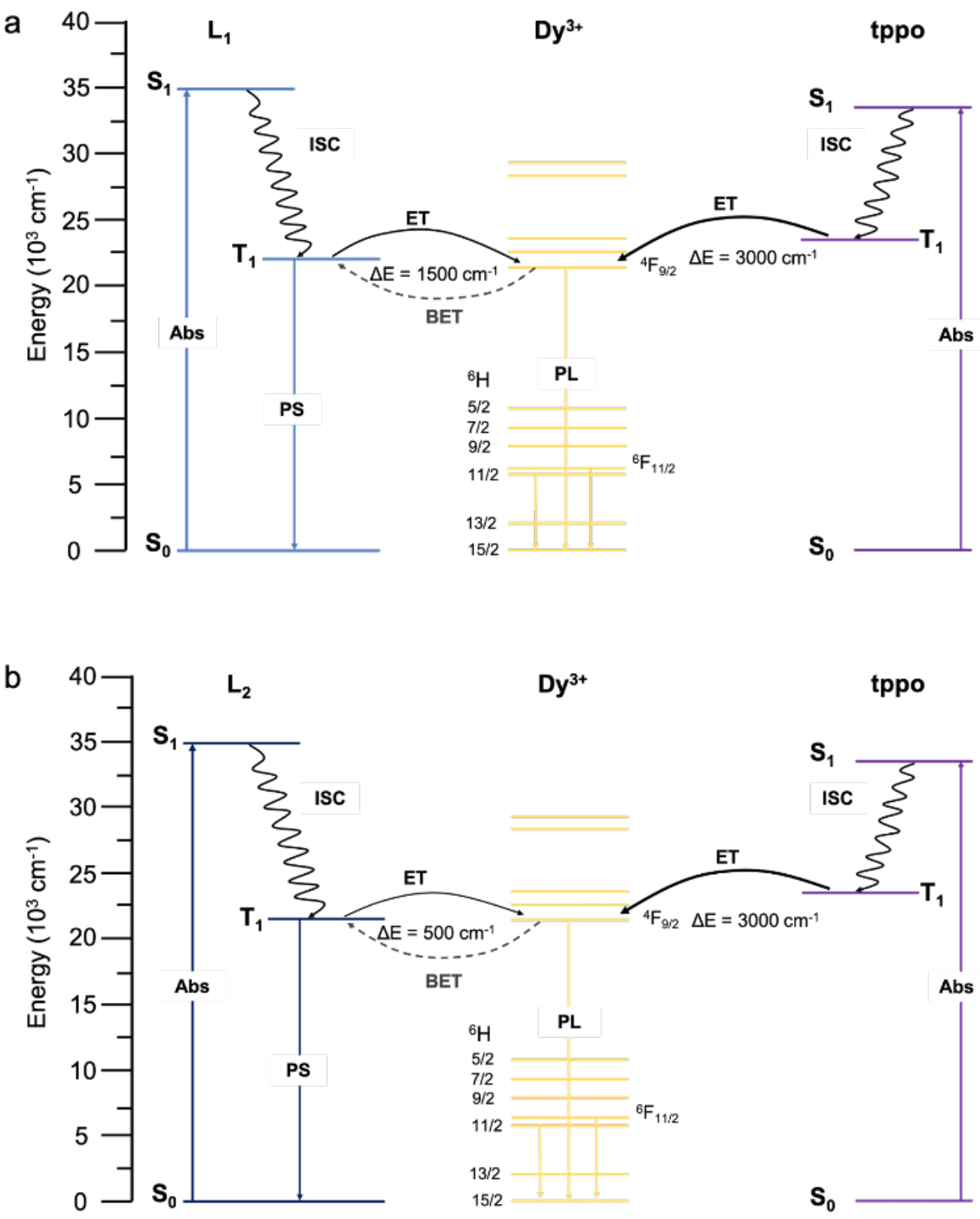

**Fig. 2.** Jablonski diagram for the $Dy^{3+}$ complexes with the ligand $L_1$ (a) and with the ligand $L_2$ (b). $S_0$ – singlet ground state, $S_1$ – singlet excited state, $T_1$ – triplet state, ISC – inter system crossing, ET – energy transfer, BET – back energy transfer, Abs – absorbance, PS – phosphorescence, PL – photoluminescence.

### 3.2.2. Steady-state and time-resolved PL of $Dy^{3+}$ complexes in the NIR region

When exciting into the absorption bands of the ligands, the investigated complexes, **$DyL_1H_2O$** and **$DyL_1$tppo** (Fig. 3a and 3b), **$DyL_2H_2O$** and **$DyL_2$tppo** (Fig. S13a and 13b in SI), show NIR luminescence. The PL spectrum in the range between 800-1650 nm is dominated by the characteristic emission peaks of $Dy^{3+}$ ion corresponding to the following transitions: ${}^4F_{9/2} \rightarrow {}^6H_{7/2} + {}^6F_{9/2}$ (846 nm), ${}^4F_{9/2} \rightarrow {}^6F_{7/2}$ (~994 nm), ${}^6F_{3/2} \rightarrow {}^6H_{13/2}$ (1066 nm), ${}^4F_{9/2} \rightarrow {}^6F_{5/2}$ (~1170 nm), ${}^6F_{11/2} \rightarrow {}^6H_{15/2}$ and ${}^6H_{9/2} \rightarrow {}^6H_{15/2}$ (~1320 nm), ${}^4F_{9/2} \rightarrow {}^6F_{1/2}$ (1404 nm) and ${}^6F_{5/2} \rightarrow {}^6H_{11/2}$ (~1500 nm). [55] However, the **$DyL_2H_2O$** complex (Fig. S13a) only shows a clear emission peak at 1317 nm while the other peaks in this region are not visible. This is likely ascribable to a significant quenching effect on the $Ln^{3+}$ ion to NIR-emission by high-strength oscillators such as C-H (ν C-H at 1650 nm and 3ν C-H 1130 nm) and O-H (2ν O-H at 1400 nm), especially for water molecules directly bonded to the emitter. [50]

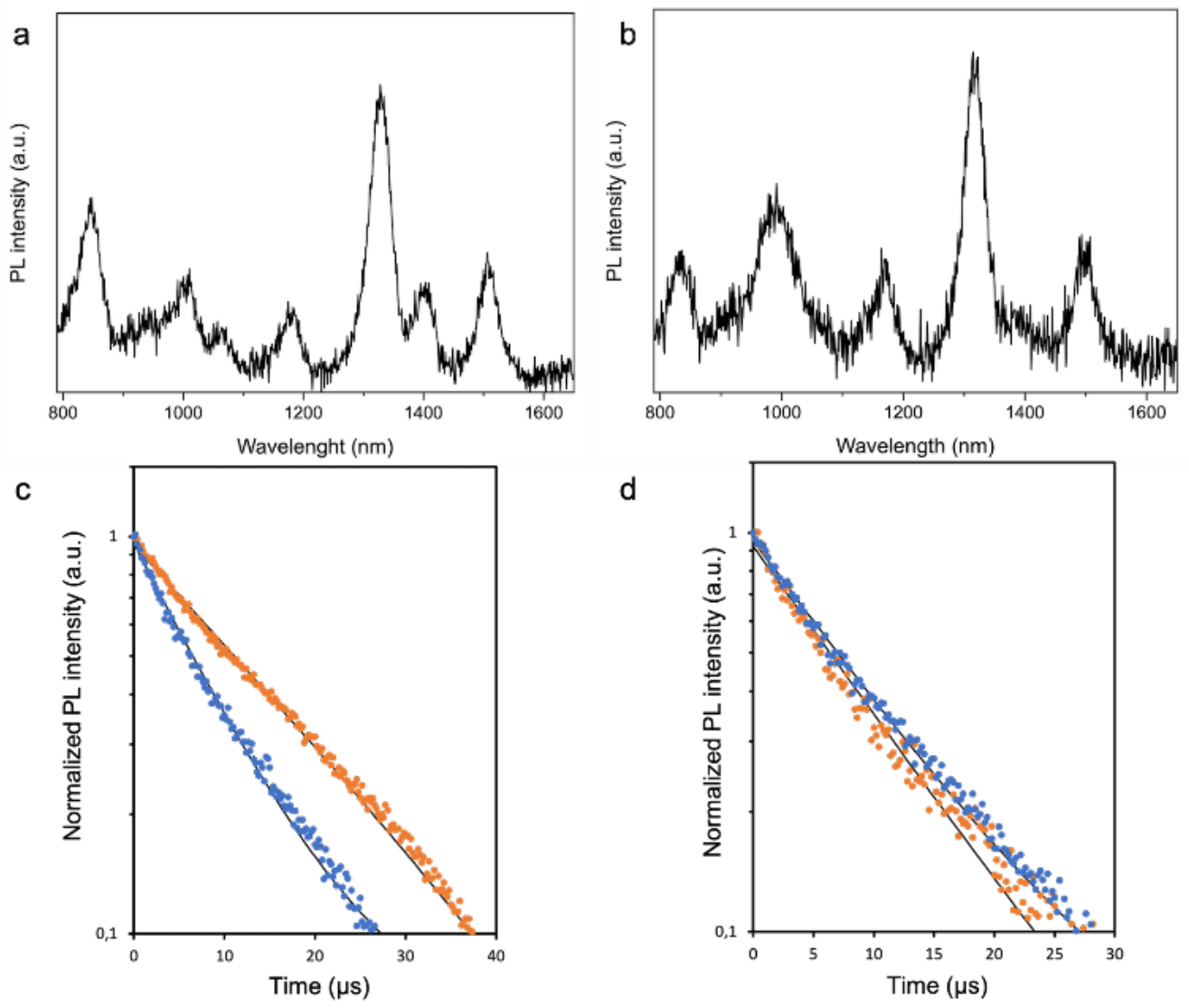

**Fig. 3.** NIR PL emission spectra of $DyL_1H_2O$ (a) and $DyL_1$tppo (b); (c) decay profile of $DyL_1H_2O$ (blue) observed at 846 nm, $DyL_1$tppo (orange) observed at 994 nm; (d) decay profile of $DyL_1H_2O$ (blue) and $DyL_1$tppo (orange) observed at 1320 nm.

The luminescence decay traces of the $Dy^{3+}$ complexes in NIR have been recorded for two peaks corresponding to the two different levels: $^4F_{9/2}$ level, monitored at 846 nm for $DyL_1H_2O$, at ~994 nm for the remaining complexes, and the $^6H_{9/2}$, $^6F_{11/2}$ levels, monitored at ~1320 nm. All decay traces are well fitted with a monoexponential function, pointing to the existence of one population of emitters and confirming the purity of the samples. As expectable, for both series of complexes with ligands $L_1$ and $L_2$, the decay profile of the $^4F_{9/2}$ $Dy^{3+}$ level reveals a significantly longer lifetime constant in the water-free derivatives than in the water-containing ones, where the emission cannot even be detected in **$DyL_2H_2O$** (Figs. 3c and S13c). This effect can be attributed to vibrational quenching through the third harmonic of the OH stretching, as was previously observed in analogous $Yb^{3+}$ complexes. [50] Interestingly, the decay dynamics of the $^6F_{11/2}$, $^6H_{9/2} \rightarrow {}^6H_{15/2}$ transitions at 1320 nm are very similar for the water-containing complexes and the tppo derivates of the same ligands. This finding seems apparently in contrast with the expected shortening of the lifetimes in the presence of bound water molecules and with observation made for the relaxation of the $^4F_{9/2}$ level. However, the change of the coordination environment, following the replacement of water molecules by tppo ligands, is likely to induce a significant variation of the oscillator strength of the transition, possibly leading to a decrease of the radiative lifetime in the water-free compounds. [50] This could therefore explain the similar observed lifetimes despite the envisaged quenching effect by water molecules. It should be also noted that the discrepancy of the observed lifetime values between the complexes with ligands $L_1$ and $L_2$ (Table 2), despite the similar amount of quenching site, is attributed to a difference in the radiative lifetime dynamics induced by the subtle differences in the ligand structure, further highlighting the relevant role of the coordination environment of the emission properties of these compounds.

**Table 2.** Observed luminescence decay time constants and corresponding $Dy^{3+}$ transitions.

| **Compound** | **λ (nm)** | **Transition** | **τ (μs)** | **$R^2$** |
|---|---|---|---|---|
| $DyL_1H_2O$ | 846 | $^4F_{9/2} \rightarrow {}^6H_{7/2} + {}^6F_{9/2}$ | 9.19 | 0.995 |
| | 1320 | $^6F_{11/2} \rightarrow {}^6H_{15/2}$ $^6H_{9/2} \rightarrow {}^6H_{15/2}$ | 10.29 | 0.995 |
| $DyL_1tppo$ | 994 | $^4F_{9/2} \rightarrow {}^6F_{7/2}$ | 17.65 | 0.997 |
| | 1320 | $^6F_{11/2} \rightarrow {}^6H_{15/2}$ $^6H_{9/2} \rightarrow {}^6H_{15/2}$ | 10.20 | 0.991 |
| $DyL_2H_2O$ | 1320 | $^6F_{11/2} \rightarrow {}^6H_{15/2}$ $^6H_{9/2} \rightarrow {}^6H_{15/2}$ | 11.99 | 0.994 |
| $DyL_2tppo$ | 994 | $^4F_{9/2} \rightarrow {}^6F_{7/2}$ | 17.37 | 0.996 |
| | 1320 | $^6F_{11/2} \rightarrow {}^6H_{15/2}$ $^6H_{9/2} \rightarrow {}^6H_{15/2}$ | 11.55 | 0.995 |

### 3.3. White-light emission (WLE) of $Dy^{3+}$ complexes

The CIE chromaticity diagram (Fig. 4.) shows that the $Dy^{3+}$ complexes with both ligands, $L_1$ and $L_2$, are emitting in the region from cold to warm white light upon excitation with UV light (365 nm). Exciting by different wavelengths in the ligand absorption bands did not result in significant emission differences. According to the CIE coordinates (Table 3) obtained for the **$DyL_1H_2O$** and **$DyL_2tppo$** complexes, the emitted white light is close to the pure white light (CIE 1931 chromaticity x = 0.333, y = 0.333) and Correlated Color Temperatures (CCT) 5470 K. Instead, the CIE coordinates for the **$DyL_1tppo$** complex are more situated toward the warm white light region leaning toward yellow-white light and the **$DyL_2H_2O$** complex shows emission in cold-white light region, corresponding to blueish-white light. As it can be seen in the CIE diagram, color tunability was achieved by changing the coordination environment by introducing the auxiliary tppo ligand. This resulted in tuning the color for the complexes with ligand $L_2$ from blue to white light, while for the complexes with ligand $L_1$ the color was tuned from white to yellow-white light.

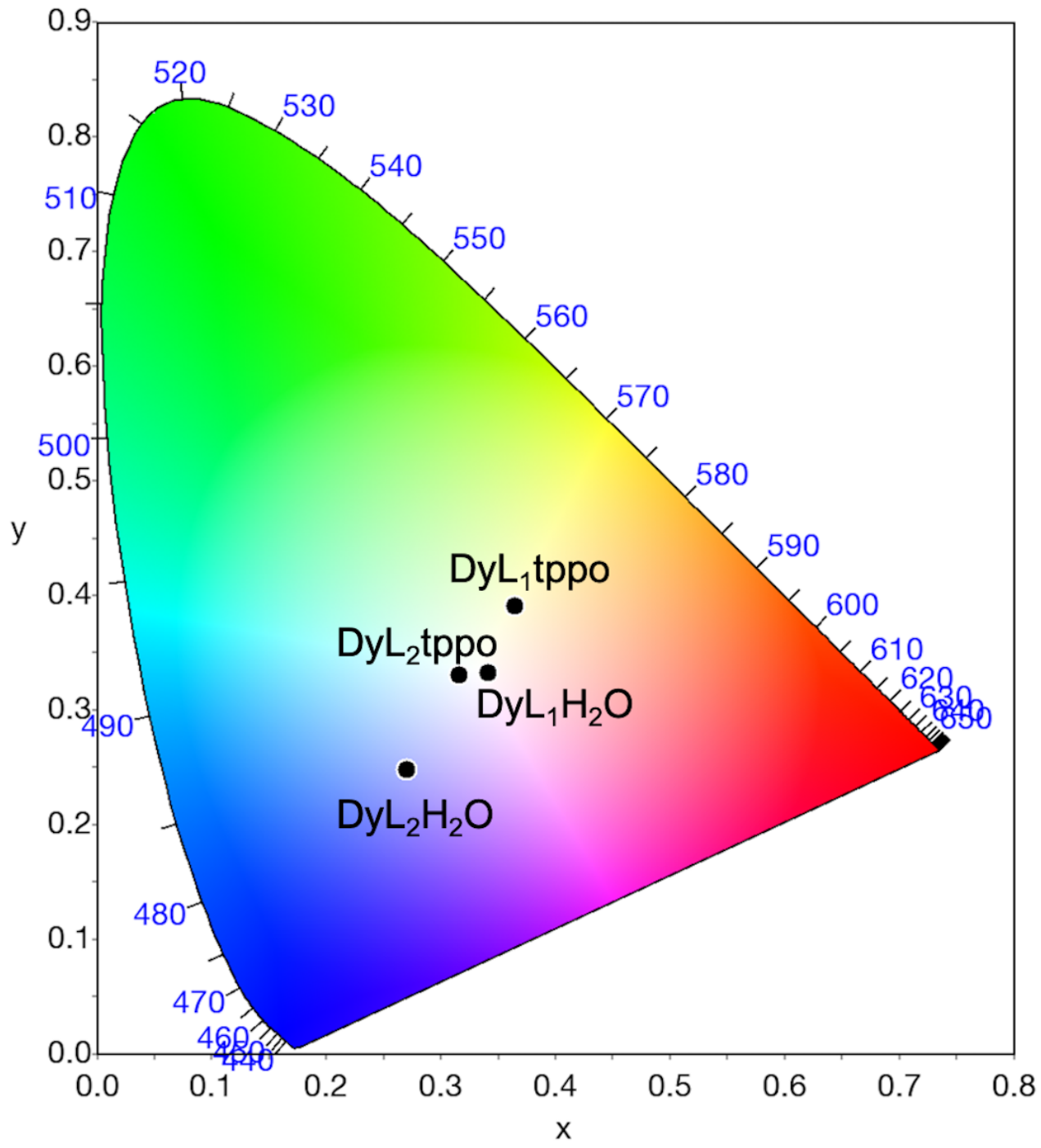

**Fig. 4.** The CIE chromaticity diagram with the color coordinates of the $Dy^{3+}$ complexes excited at 365 nm.

**Table 3.** CIE color coordinates (x,y) and CCT for the $Dy^{3+}$ complexes in the solid state.

| **Compound** | **x** | **y** | **CCT (K)** |
|---|---|---|---|
| $DyL_1H_2O$ | 0.340 | 0.333 | 5129 |
| $DyL_1$tppo | 0.364 | 0.391 | 4537 |
| $DyL_2H_2O$ | 0.270 | 0.249 | 18173 |
| $DyL_2$tppo | 0.316 | 0.331 | 6319 |

## 4. Conclusion

A series of water-containing and water-free $Dy^{3+}$ β-diketonate complexes were prepared with two ligands, $L_1$ and $L_2$, with similar chemical properties but with a slight structural difference. In the phenyl ring of ligand $L_2$ the H-atom in *para*-position to the β-diketonate group was replaced with a Cl-atom. The water-free complexes have been prepared with a neural tppo co-ligand that excludes water molecules from the first coordination sphere of the $Dy^{3+}$ ion and acts as an additional sensitizer for energy transfer to the nearby $Dy^{3+}$ ion. All the complexes showed emission with characteristic transitions in both Vis and NIR region. The proximity of the energies of the triplet ($T_1$) states of ligands $L_1$ and $L_2$ to the emitting level of $Dy^{3+}$ ($^4F_{9/2}$) led to the observation of ligand emission in addition to the emission of the $Dy^{3+}$ ion. This was exploited to obtain white-light emission. The observed WLE of **$DyL_1H_2O$** (x = 0.340, y = 0.333) and **$DyL_2$tppo** (0.316, 0.331) was close to the idea white light with the CCT in the clod white-light region. Interestingly, the complexes also rarely observed $Dy^{3+}$ - centered NIR emission with a peak at 1320 nm falling in the O-band region of interest in optical telecommunications. Careful design of a ligand that would promote the emission in the NIR region, without enhancing the quenching of $Dy^{3+}$ emission in this region, could lead to use of $Dy^{3+}$ as an alternative to the well-know lanthanide ions, such as $Er^{3+}$ and $Nd^{3+}$, that are currently used for optical telecommunications.

## Conflict of interest

There are no conflicts to declare.

## Acknowledgment

DM and TV acknowledge KU Leuven Postdoctoral Mandate Internal Funds (PDM) for a postdoctoral fellowship (PDM/20/092). FA acknowledges the FWO (Research Foundation Flanders) and the European Union's Horizon 2020 program for a [Pegasus]$^2$ Marie Skłodowska-Curie fellowship "Nanocomposite materials for highly efficient sensitized lanthanide emission". BK and BB acknowledge financial support of The Ministry of Education, Science and Technological Development of The Republic of Serbia. BK also acknowledge support from F R S-FNRS.

## Supporting Information

In the Supporting Information are the additional PL spectra, FT-IR and ESI-MS data.